# Magnetic Activity of F-, G-, and K-type Stars in the LAMOST–*Kepler* Field

Jinghua Zhang[1], Shaolan Bi[1,10], Yaguang Li[1], Jie Jiang[2], Tanda Li[3,4], Han He[5], Jie Yu[6], Shourya Khanna[3], Zhishuai Ge[7], Kang Liu[1], Zhijia Tian[8], Yaqian Wu[9], and Xianfei Zhang[1]
[1] Department of Astronomy, Beijing Normal University, Beijing 100875, People's Republic of China; zhangjinghua@mail.bnu.edu.cn, bisl@bnu.edu.cn
[2] School of Space and Environment, Beihang University, Beijing 100083, People's Republic of China
[3] Sydney Institute for Astronomy (SIfA), School of Physics, University of Sydney, Sydney, NSW 2006, Australia
[4] Stellar Astrophysics Centre, Department of Physics and Astronomy, Aarhus University, Ny Munkegade 120, DK-8000 Aarhus C, Denmark
[5] CAS Key Laboratory of Solar Activity, National Astronomical Observatories, Chinese Academy of Sciences, Beijing 100101, People's Republic of China
[6] Max Planck Institute for Solar System Research, D-37077 Göttingen, Germany
[7] Key Laboratory of Beam Technology of Ministry of Education, Beijing Radiation Center, Beijing 100875, People's Republic of China
[8] School of Physics and Astronomy, Yunnan University, Kunming 650091, People's Republic of China
[9] Key Laboratory of Optical Astronomy, National Astronomical Observatories, Chinese Academy of Sciences, Beijing 100101, People's Republic of China
Received 2019 April 30; revised 2019 December 2; accepted 2019 December 10; published 2020 February 20

## Abstract

Monitoring chromospheric and photospheric indexes of magnetic activity can provide valuable information, especially the interaction between different parts of the atmosphere and their response to magnetic fields. We extract chromospheric indexes, $S$ and $R_{HK}^+$, for 59,816 stars from LAMOST spectra in the LAMOST–*Kepler* program, and photospheric index, $R_{\rm eff}$, for 5575 stars from *Kepler* light curves. The $\log R_{\rm eff}$ shows positive correlation with $\log R_{HK}^+$. We estimate the power-law indexes between $R_{\rm eff}$ and $R_{HK}^+$ for F-, G-, and K-type stars, respectively. We also confirm the dependence of both chromospheric and photospheric activity on stellar rotation. Ca II $H$ and $K$ emissions and photospheric variations generally decrease with increasing rotation periods for stars with rotation periods exceeding a few days. The power-law indexes in exponential decay regimes show different characteristics in the two activity–rotation relations. The updated largest sample including the activity proxies and reported rotation periods provides more information to understand the magnetic activity for cool stars.

*Unified Astronomy Thesaurus concepts:* Stellar activity (1580); Stellar atmospheres (1584); Stellar chromospheres (230); Stellar magnetic fields (1610)

*Supporting material:* machine-readable table

## 1. Introduction

The study of stellar activity in terms of different emission features originating from different parts of atmospheres has demonstrated the close connection between stellar activity, rotation, magnetic field, and inner stellar dynamo (Babcock 1958; Middelkoop 1982; Noyes et al. 1984). The magnetic fields generated through magnetohydrodynamic processes control the structure and the energetic balance of stellar atmospheric plasma, bringing a variety of phenomena, e.g., inhomogeneous dark spots and bright faculae from the photosphere, emissions in the line cores from the chromosphere and the transition region, and thermal X-rays and eruptive flares from the coronal region (Catalano et al. 1999).

The chromospheric Ca II $H$ and $K$ emissions are well known to correlate well with the stellar actual magnetic flux and thus are the most commonly used indicator of chromospheric activity for cool stars (Saar & Schrijver 1987; Chatzistergos et al. 2019). The Mount Wilson program measured the chromospheric Ca II $H$ and $K$ emissions of more than 1000 stars for over four decades (Wilson 1963; Duncan et al. 1991; Baliunas et al. 1995). These emissions were quantified as the Mount Wilson $S$ value ($S_{\rm MW}$; Duncan et al. 1991). It was used to monitor long-term stellar chromospheric activity (Baliunas et al. 1995; Lockwood et al. 2007). However, there are some photospheric flux contributions in the wings of the $H$ and $K$ lines so that the values of $S_{\rm MW}$ do not depend on chromosphere solely (Linsky & Ayres 1978). Noyes et al. (1984) derived the chromospheric emission fraction parameter $R'_{HK}$, which was converted from $S$ by removing an empirically determined photospheric contribution $R_{\rm phot}$. Schrijver (1987) introduced the concept of basal components in the chromospheric flux emissions from an atmosphere heated by acoustic waves and shocks. The basal flux characterizes stars with minimal activity levels and depends sensitively on effective temperature (Schrijver et al. 1989; Rutten & Uitenbroek 1991). Using the $S$ index from large samples with different luminosity classes, Mittag et al. (2013) parameterized the empirical basal chromospheric flux. They derived a new activity indicator $R_{HK}^+$ to characterize the pure activity-related Ca II $H$ and $K$ line surface flux for stars of different spectral types.

The unprecedented quality of the continuous four-year photometric observations carried out by the *Kepler* space mission have extended our understanding of photospheric activity to a large number of field stars (Borucki et al. 2010). Its broad optical photometry measures the variability caused by starspots or faculae zone rotating into and out of visibility as the star rotates. Hence, the range or amplitude of light-curve fluctuation is commonly used as a proxy for stellar photospheric activity (e.g., Basri et al. 2013; García et al. 2014). Basri et al. (2013) used the range between the 5th and 95th percentile of flux as a proxy for photometric variability. This method could underestimate the variability of very active stars (García et al. 2014). Instead, García et al. (2014) defined a new index of photometric variability ($S_{\rm ph}$) as the mean value of the light-curve fluctuations over subseries of length $5 \times P_{\rm rot}$, where $P_{\rm rot}$ is the rotation period of the star. Since the fluctuation amplitude is not generally uniform for a given stellar light curve, a reasonable quantity ($R_{\mathrm{eff}}$) was defined by He et al. (2015) to represent the effective range of the light-curve fluctuation amplitude.

---
[10] Corresponding author.

It might be expected that more rapid rotation would lead to either an increased number of spots and/or larger spots overall, similar to the patterns between rotation and chromospheric activities. Therefore, the amplitude of light-curve fluctuation and the chromospheric activities tend to be related to each other. Notsu et al. (2015) investigated the connection between mean stellar brightness variation and the residual flux in the infrared Ca II line core, and showed that the two quantities are strongly correlated. Karoff et al. (2016) confirmed the correlation for 1400 G-type stars and found that such correlation is absent for stars with activity levels lower than the Sun. There is extensive research on the relations between different activity proxies for the solar case (e.g., Bennett et al. 1984; Cappelli et al. 1989; Schrijver et al. 1989). Schrijver et al. (1989) originally derived a power-law index of approximately 0.6 between the Ca II K-line core excess flux density and the absolute value of the magnetic flux density for solar active regions. A power-law exponent of 0.2 was suggested by Rezaei et al. (2007) for locations in a quiet Sun and higher values of 0.4–0.5 for network locations. Schrijver et al. (1989) interpreted the below-unity power-law exponent as the geometric expansion model of magnetic flux tubes. This qualitative picture was confirmed by Solanki et al. (1991) by applying a two-dimensional magnetostatic model. Recently, Barczynski et al. (2018) used the original geometric expansion model to explain the relations between emissions of solar chromospheric or transition regions and the magnetic fields. Whether correlations between photospheric and chromospheric activity exist as seen in a large range of spectral types is still an open question. The Large Sky Area Multi-Object Fibre Spectroscopic Telescope (LAMOST) spectroscopic survey (Cui et al. 2012; Zhao et al. 2012) has collected millions of stellar spectra with a resolution of 1800 in a broad wavelength range of 3700–9000 Å. This wealth of data provides a golden opportunity to answer this question.

Stellar rotation plays an important role in the dynamo and affects magnetic activity. The rotational spin-down due to the loss of stellar angular momentum weakens the efficiency of the dynamo, leading to a decreasing magnetic activity (Parker 1955; Skumanich 1972). The relationships between stellar rotation period and the magnetic activity levels in terms of chromospheric and coronal proxies have been studied. The relations show different configurations, with fast rotators falling in the saturated activity regime, and slow rotators falling in the exponential decay regime. (e.g., Kraft 1967; Noyes et al. 1984; Pizzolato et al. 2003; Wright et al. 2011). In addition, the relations between rotation period and photospheric activity proxies have been investigated (McQuillan et al. 2014). With the help of LAMOST and the *Kepler* mission, we can study the relations among stellar rotation periods, the chromospheric activity proxies, and the photospheric activity proxies in a large sample comprehensively.

In this paper, we construct a large sample of cool stars with photospheric and chromospheric activity proxies in the LAMOST–*Kepler* field. Section 2 describes methods of data analysis. We explore the relations between the activity indexes in Section 3. In Section 4, we investigate the activity–rotation relationship in detail. We summarize the conclusions in Section 5.

## 2. Data Analysis

The LAMOST–*Kepler* project was initiated to use the LAMOST spectroscopic survey to perform spectroscopic follow-up observations for the targets in the field of the *Kepler* mission (De Cat et al. 2015). By 2016 June, this project had collected more than 180,000 optical spectra covering 3700–9000 Å in low-resolution $R \sim 1800$. In this work, we selected targets from the LAMOST DR4. The spectra were processed by the LAMOST Spectroscopic Survey of Galactic Anticentre (LSS–GAC) flux calibration pipeline (Liu et al. 2014; Yuan et al. 2014). We use stellar atmospheric parameters ($T_{\mathrm{eff}}$, $\log g$, and [Fe/H]) from the LAMOST Stellar Parameter Pipeline at Peking University (LSP3; Xiang et al. 2015). For the spectra with signal-to-noise ratios (S/Ns) higher than 50, stellar parameters from LSP3 have typical uncertainties of 100 K for $T_{\mathrm{eff}}$, 0.1 dex for $\log g$, and 0.1 dex for [Fe/H] (Xiang et al. 2017).

We aim at targets of dwarfs, removing giant stars with an empirical $T_{\mathrm{eff}}$–$\log g$ relation determined by Ciardi et al. (2011). Binaries labeled by Berger et al. (2018) were also excluded. To select F- to K-type stars, we use $T_{\mathrm{eff}}$ in the range of 3800–7200 K. To place a lower limit on the quality of the individual observations, the S/Ns at the blue end of the spectra are higher than 10. With these constraints, we gathered 86,689 spectra for 59,816 stars.

For photospheric activity analysis, we selected the sample stars from catalog in McQuillan et al. (2014). Note that this catalog is biased to stars that produce measurable rotational curves, which leaves out photometrically quiet or long-period stars. This catalog provides the updated largest sample set with rotation period. The period is acquired through rotational modulation due to the existence of inhomogeneous spots and faculae that lead to fluctuating light curves. We cross match 59,816 stars with the catalog of McQuillan et al. (2014) and obtain 5575 targets with both photometric observational data and spectroscopic observational data.

### 2.1. Quantifying Chromospheric Activity

#### 2.1.1. Determining the S Index

The chromospheric activity level is typically quantified through the classical $S$ index, i.e., the ratio of the flux in the core of the Ca II $H$ and $K$ lines to the nearby continuous windows (Vaughan et al. 1978). Figure 1 shows examples of spectra for typical FGK-type stars at different activity levels. The spectra of these stars have been normalized in the spectral range of 3900 Å–$\sim$4000 Å. Following Karoff et al. (2016, hereafter K16), we computed the flux ratio $S$ as the emission in the Ca II $H$ and $K$ lines relative to the continuum,

$$S = 8\alpha \cdot \frac{H + K}{R + V}, \quad (1)$$

where $H$ and $K$ are the fluxes integrated in 1.09 Å FWHM triangular windows centered on the line cores of 3968 and 3934 Å. $R$ and $V$ are the fluxes integrated in 20 Å rectangular windows centered on 4001 Å and 3901 Å. The normalization factor $\alpha = 1.8$ was adopted from Hall et al. (2007). The factor of 8 is the ratio of exposure time between $HK$ and $RV$ channels of the Mount Wilson HKP-2 spectrophotometer. For stars with multiple observations, the $S$ values were calculated by the weighted mean values of these multiple spectra.

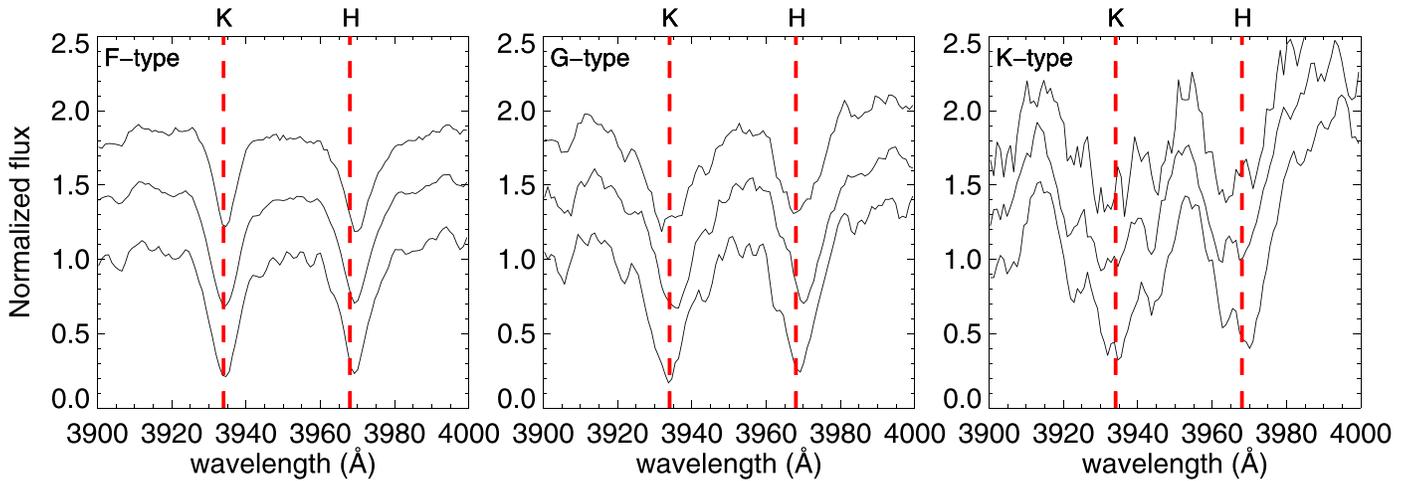

**Figure 1.** Representative spectra of different chromospheric emission levels in F-, G-, and K-type stars. A vertical offset of 0.4 is applied between each spectrum for clarity. The spectra lines from bottom to top in each panel are the stars with different emission levels. The cores of Ca II *H* and *K* emission lines are indicated by red dashed lines in each panel.

To estimate the uncertainties of *S* indexes, we applied a similar procedure as in K16. We calculated the relation of the standard deviation of different S measurements as a function of the mean S/N at the spectra's blue end for stars with multiple observations as $\log \sigma(S) = -1.5 \log(S/N) + 1.0$. The uncertainties given by this relation are considered the causes of stellar intrinsic chromospheric activity variation. We also considered the random errors. We used a Monte Carlo approach to estimate them. We added Gaussian noises to the original spectrum to generate a simulated spectrum. The *S* value was then calculated for the simulated spectrum. This was done 1000 times, and the standard deviation of the 1000 *S* values was adopted as the uncertainty. We finally took the standard error of the uncertainties given by the two procedures as the *S* measurement uncertainty for each star.

K16 calculated the *S* index for ∼4000 G-type stars in the LAMOST–*Kepler* field. In Figure 2 we compare our *S* index values with those of K16 for stars in common. There is a good agreement between the two sets of *S* values. The difference may arise from the fact that we used spectra processed with the LSP3 pipeline, while K16 used spectra processed with the LAMOST Stellar Parameter pipeline (LASP; Luo et al. 2015). We also compared the *S* index distribution with that of Isaacson & Fischer (2010) and found agreement for moderately active stars but a lack of high-activity stars in the LAMOST sample. Given the low-resolution power of LAMOST, the wavelength ranges of the *HK* emissions are not easy to identify. Thus, it is possible to take some fluxes outside the veritable emission windows in the S measurements, which would lead to different results between high-resolution spectra and low-resolution spectra. K16 found a similar lack of high-activity stars in the LAMOST sample as well.

### 2.1.2. Determining the $R_{HK}^+$ Index

The quantity *S* is sensitive to the integrated emission over these windows and the photospheric radiation transmitted by *H* and *K* instrumental passbands, both of which are color dependent (Middelkoop 1982). As mentioned in Section 1, Mittag et al. (2013, hereafter M13) defined a new proxy, $R_{HK}^+$, which is converted from the *S* index by eliminating the photospheric contributions (Noyes et al. 1984) and the so-called basal chromospheric flux (Schrijver 1987). Using M13's

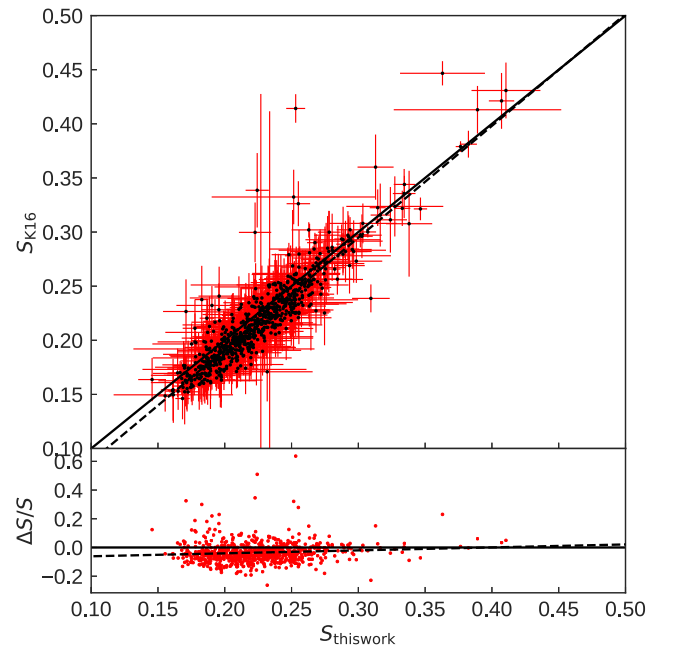

**Figure 2.** Comparison of the *S* index with those of Karoff et al. (2016) for stars in common. The solid line shows the line of equality; the dashed line shows a least-squares method fitting of the results. In the lower panel, the dispersions for the *S* index are plotted on the *Y*-axis.

method, we calculated the index $R_{HK}^+$ following

$$R_{HK}^+ = \frac{\mathcal{F}_{HK} - \mathcal{F}_{HK,\mathrm{phot}} - \mathcal{F}_{HK,\mathrm{basal}}}{\sigma T_{\mathrm{eff}}^4} = \frac{\mathcal{F}_{HK}^+}{\sigma T_{\mathrm{eff}}^4}, \quad (2)$$

where $T_{\mathrm{eff}}$ is the effective temperature, and $\sigma$ is the Stefan-Boltzmann constant. Here the *HK* surface flux is derived from the *RV* continuum flux and the *S* index through $\mathcal{F}_{HK} = S \cdot \mathcal{F}_{RV}/(8\alpha)$ (Middelkoop 1982). The photospheric flux in the *HK* bands $\mathcal{F}_{HK,\mathrm{phot}}$, the basal chromospheric flux $\mathcal{F}_{HK,\mathrm{basal}}$, and the continuum flux $\mathcal{F}_{RV}$ were calculated from the *B*−*V* color index (Ramírez & Meléndez 2005). The uncertainties of $R_{HK}^+$ were estimated by considering the uncertainties of the *S* index. Note that both $\mathcal{F}_{HK,\mathrm{phot}}$ and $\mathcal{F}_{HK,\mathrm{basal}}$ are virtually given by the empirical formula that is based on stellar color.

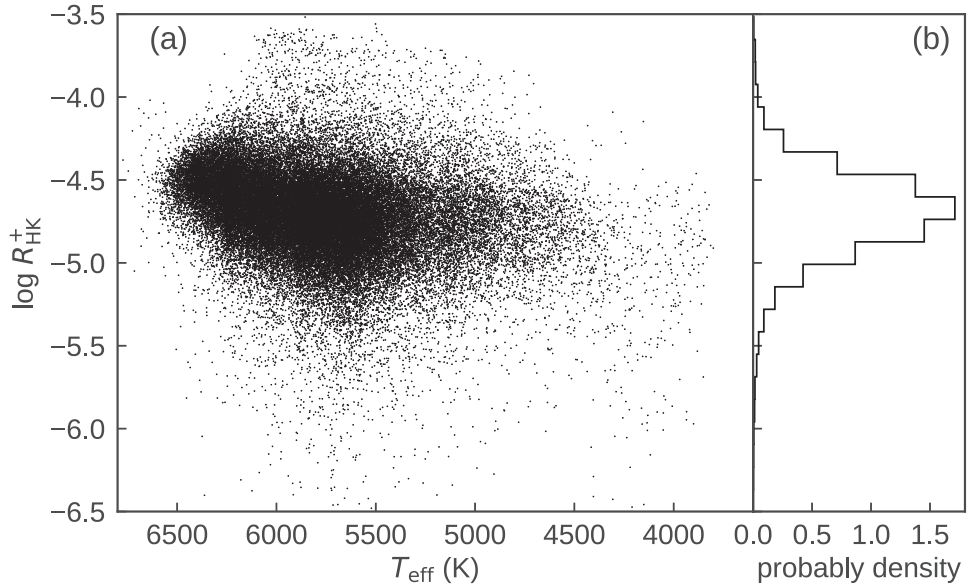

**Figure 3.** Distribution of $\log R_{HK}^+$ with $T_{\rm eff}$ of all the dwarfs that have the chromospheric observations. Panel (b) shows the histogram of $\log R_{HK}^+$.

**Table 1**
Sample Entries of Deduced Activity Proxies of the 5575 Stars

| KIC | $T_{\rm eff}$ | $\log g$ | [Fe/H] | $S$ | $\log R_{HK}^+$ | $R_{\rm eff}$ |
|---|---|---|---|---|---|---|
| 1028018 | 5436.00 ± 79.74 | 4.14 ± 0.11 | 0.03 ± 0.11 | 0.3521 ± 0.0175 | −4.3130 ± 0.0291 | 0.058877 ± 0.003468 |
| 1161620 | 5752.83 ± 131.96 | 4.74 ± 0.17 | 0.04 ± 0.11 | 0.2654 ± 0.0303 | −4.4721 ± 0.0826 | 0.011962 ± 0.000984 |
| 1163579 | 5443.16 ± 71.63 | 4.44 ± 0.09 | −0.15 ± 0.09 | 0.2675 ± 0.0099 | −4.4796 ± 0.0697 | 0.012592 ± 0.001091 |
| 1292666 | 5517.19 ± 66.86 | 4.50 ± 0.08 | 0.31 ± 0.08 | 0.1869 ± 0.0070 | −4.9468 ± 0.0312 | 0.004682 ± 0.000270 |
| 1295597 | 5708.76 ± 69.91 | 4.41 ± 0.12 | −0.15 ± 0.06 | 0.2224 ± 0.0056 | −4.6105 ± 0.0198 | 0.003783 ± 0.000242 |
| ... | ... | ... | ... | ... | ... | ... |

**Note.** Columns 2–4 are the parameters from Xiang et al. (2015). Stars without $\log R_{HK}^+$ are labeled with NaN.

(This table is available in its entirety in machine-readable form.)

The color range of the sample selected in M13 is $0.44 < B - V < 1.6$. For early F-type stars ($B - V < 0.44$), the $R_{HK}^+$ values calculated by Equation (2) are not reasonable. Figure 3 shows the distribution of $\log R_{HK}^+$ with $T_{\rm eff}$ of our sample. The histogram of $\log R_{HK}^+$ is in the right panel. The values of $\log R_{HK}^+$ for most stars are in the range of $-4.3 \sim -5.0$. For K-type stars, the values of $\log R_{HK}^+$ are mostly lower than $\sim -4.6$.

### 2.2. Quantifying Photospheric Activity

The photometric data were obtained by the *Kepler* mission and were acquired in the long-cadence mode (29.4 minutes; Jenkins et al. 2010). We used data processed by the Presearch Data Conditioning module (Smith et al. 2012; Stumpe et al. 2012) of the *Kepler* data analysis pipeline.

We used $R_{\rm eff}$ to represent the photospheric activity (He et al. 2015, 2018). For a given light curve $F_t$, $t = 0, 1, 2, 3, ..., N-1$ with $N$ points, we first obtain the relative flux:

$$f_t = \frac{F_t - \widetilde{F}}{\widetilde{F}}, \quad (3)$$

where $\widetilde{F}$ is the median of $F_t$. Then, we used a Fourier-based low-pass filter to remove high-frequency variations present in $f_t$. The nature of these variations could be outliers, flare spikes, oscillation signals, and granulation-driven flickers (Cranmer et al. 2014; Kallinger et al. 2014). The cutoff frequency is given by the empirical relation, $f_{\rm upper} = \frac{1}{0.3 P_{\rm rot}}$, which takes into account that different stars may have different noise levels and fluctuation properties (He et al. 2015; Mehrabi et al. 2017). Finally, we obtained the pure gradual variation component, $f_G$. The effective fluctuation range of the light curve is given by

$$R_{\rm eff} = 2 \cdot (\sqrt{2} \cdot f_{\rm rms}), \quad (4)$$

where $f_{\rm rms}$ is the rms value of $f_G$ (García et al. 2010; Chaplin et al. 2011). The factor $2\sqrt{2}$ in Equation (4) is given to introduce a corrected value of the fluctuation range; see He et al. (2015) for a detailed illustration. For each star, we calculated $R_{\rm eff}$ quarter by quarter and then took the average value as the evaluated proxy. The uncertainty of $R_{\rm eff}$ was taken by the standard error of $R_{\rm eff}$ values in all quarters. Table 1 lists $T_{\rm eff}$, $\log g$, [Fe/H], $S$, $\log R_{HK}^+$, and $R_{\rm eff}$ of our sample.

In Figure 4, we plot the distributions of $\log R_{\rm eff}$ with $T_{\rm eff}$. The histogram of $\log R_{\rm eff}$ is plotted in the right panel. The difference of photospheric activity levels is almost 3 orders of magnitude between the most active stars and the inactive stars. The photospheric activity levels ($\log R_{\rm eff}$) for most stars are in the range of $-2.7 \sim -1.8$. The average $\log R_{\rm eff}$ values of each temperature bin shown by the red line increase with decreasing $T_{\rm eff}$. When $T_{\rm eff} < 5300$ K, the $\log R_{\rm eff}$ values are around $-2.3 \sim -2.2$. This is consistent with the result of Giles et al. (2017), who took a similar photospheric activity index to investigate the correlation between the size of starspots and the stellar effective temperature.

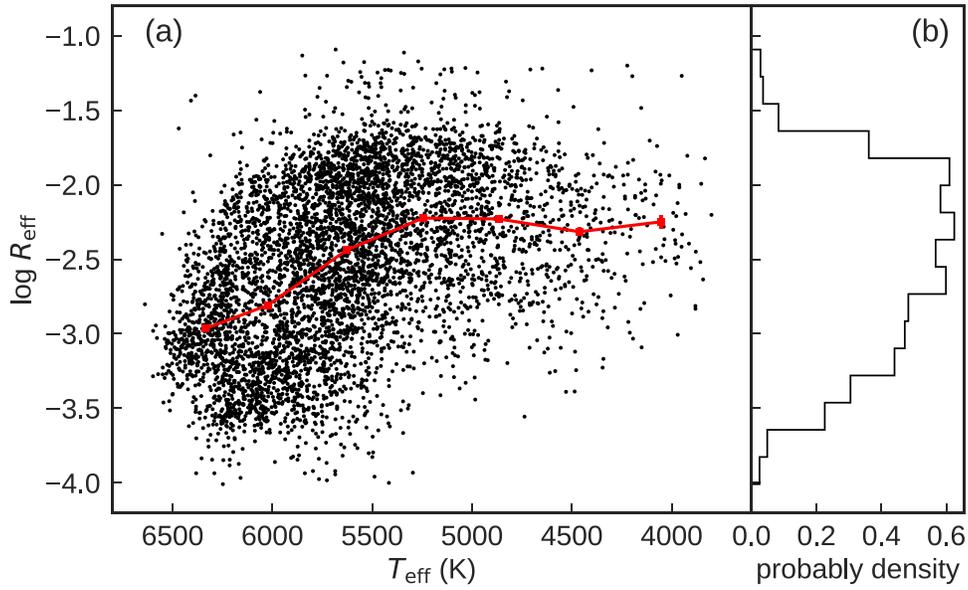

**Figure 4.** Distribution of $\log R_{\rm eff}$ with $T_{\rm eff}$ of our sample. The average $\log R_{\rm eff}$ values of each temperature bin are shown by the red line. The sample stars are divided into seven bins by $T_{\rm eff}$. Panel (b) shows the histogram of $\log R_{\rm eff}$.

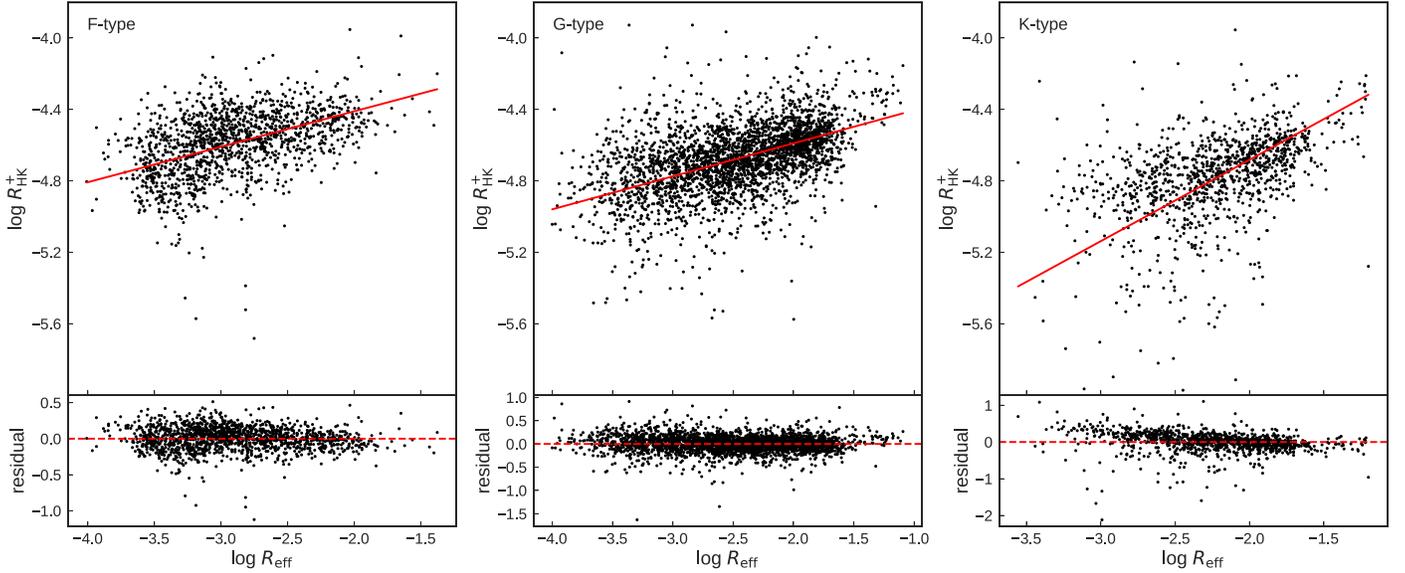

**Figure 5.** Relations between logarithmic $R_{\rm eff}$ and $R^{+}_{HK}$ index for F-type (left), G-type (middle), and K-type (right) stars. The red lines represent the power-law approximation. The residuals are plotted on the Y-axis in the respective lower panels.

### 3. Relations between Chromospheric and Photospheric Activities

We compared $R_{\rm eff}$ with $R^{+}_{HK}$ for F-, G-, and K-type stars in Figure 5. The numbers of F-, G-, and K-type stars are 1444, 2997, and 1134, respectively.

As shown in Figure 5, for F-, G-, and K-type stars, the $\log R^{+}_{HK}$ values show positive correlations with the $\log R_{\rm eff}$ values. We use Spearman's rank order correlation coefficient ($r_s$) to test the potential connections between the two quantities. The $r_s$ values for the F-, G-, and K-type stars are 0.49, 0.50, and 0.53, respectively, indicating that there is a relation between $R_{\rm eff}$ and $R^{+}_{HK}$. The scatter is probably due to the physical differences between measuring photometric intensity contrasts and chromospheric emissions. The different observational time lengths between photometric and spectroscopic observations may be another reason. For each star, the $R_{\rm eff}$ was obtained from continuous observation in 4 yr, while the $R^{+}_{HK}$ was obtained from a few observational records. To estimate these relations, we performed an orthogonal regression method (Isobe et al. 1990; Feigelson & Baru 1992) in terms of

$$\log R^{+}_{HK} = c + k \cdot \log R_{\rm eff}, \quad (5)$$

where $c$ is a scaling parameter and $k$ is a power-law index. The fitted power-law indexes are listed in Table 2 with fitted relations shown in Figure 5 by red lines. The power-law exponents for F- and G-type stars are similar, while for K-type stars the exponent is larger than that of earlier type stars. The reason for this difference may be also related to the lack of high-activity stars at the cool end in our sample.

**Table 2**
Power-law Indexes Fit for Different Type Stars in This Work

| Index | F-type | G-type | K-type |
|---|---|---|---|
| k | 0.20 ± 0.08 | 0.18 ± 0.04 | 0.46 ± 0.09 |
| c | −4.01 ± 0.22 | −4.22 ± 0.10 | −3.77 ± 0.20 |
| rms | 0.16 | 0.17 | 0.27 |

The relationship between chromospheric activity and photospheric variability is also investigated on the decadal timescale of the solar activity cycle (Radick et al. 1998; Lockwood et al. 2007). Lockwood et al. (2007) showed that on a year-to-year timescale, the young active stars become fainter as their Ca II $H$ and $K$ emission increases, and older less active stars tend to show a pattern of direct correlation. The correlation studied in this work is based on the stellar rotational timescale. The results indicate that on rotational timescales, the photospheric variability always shows positive correlation with the Ca II $H$ and $K$ emissions for cool stars. On the other hand, the relations between the chromosphere activity proxies and the activity proxies in transition region or in the coronal region have also been studied in the stellar case (e.g., Oranje 1986; Schrijver et al. 1992; Güdel 2004). In this work, we additionally give the relation between activity proxies in the photosphere and in the chromosphere. Our results and the previous results provide related magnetic information from the photosphere to coronal regions for cool stars.

## 4. Activity–Rotation Relations

We studied the relations between stellar rotation periods and chromospheric emission as well as photospheric variation for F-, G-, and K-type stars. Figure 6 and Figure 7 show $R_{HK}^+$ versus $P_{\rm rot}$ and $R_{\rm eff}$ versus $P_{\rm rot}$. The histograms of rotation periods for each subset are shown in the respective upper panels. We found that when the rotation periods exceed about 1, 3, and 6 days for F-, G-, and K-type stars, the probabilities of the density distribution become significant. Beyond that, the corresponding rotation periods at the density peaks are near 6, 18, and 25 days, which become longer from F- to K-type stars.

Compared with K-type stars, the activity–rotation relations are much more dispersive for F- and G-type stars, especially in the $R_{\rm eff}$–$P_{\rm rot}$ relation. This phenomenon is also observed in the flare and X-ray band (Huiqin & Jifeng 2019; Pizzolato et al. 2003). For stars with rotation periods exceeding about 1, 3, and 6 days, Ca II $H$ and $K$ emissions and photospheric variations decrease significantly with increasing rotation periods, forming an exponential decay regime.

We parameterized the relations for F-, G-, and K-type stars that possess $P_{\rm rot}$ larger than 1, 3, and 6 days, individually. We fitted the activity index ($R_{HK}^+$ and $R_{\rm eff}$) versus $P_{\rm rot}$ with a power law as the form:

$$\log R_{\rm i} = c + \beta \cdot \log P_{\rm rot}, \quad (6)$$

where $R_{\rm i}$ is $R_{HK}^+$ or $R_{\rm eff}$, $c$ is the scaling factor, and $\beta$ is the power-law index (Wright et al. 2011). The parameters were determined by the Ordinary Least Squares bisector (Isobe et al. 1990; Feigelson & Baru 1992). The fitted results with respective errors are listed in Table 3. In $R_{HK}^+$–$P_{\rm rot}$ relations, the absolute values of $\beta$ increase from F- to K-type stars, which means the slope of the exponential decay regime becomes steeper. In $R_{\rm eff}$–$P_{\rm rot}$ relations, the $\beta$ value of F-type stars is less than that of G- and K-type stars. Beyond that, the slopes in $R_{\rm eff}$–$P_{\rm rot}$ relations are all steeper than those in $R_{HK}^+$–$P_{\rm rot}$ relations, which implies that the dependence of the photometric intensity contrasts on stellar rotation is different from the dependence of the chromosphric emissions on rotation.

The similar decreasing trends of activity proxies with increasing rotation periods have also been shown in previous works (Pizzolato et al. 2003; Wright et al. 2011; Wright & Drake 2016). The explanation for this trend might be based on the $\alpha\omega$-type dynamo theory (Noyes et al. 1984; Charbonneau 2014). That is, the observed decrease in proxies of stellar activity driven by the stellar magnetic dynamo could be attributed to the rotational spin-down of the star, which is driven by mass loss through a magnetized stellar wind (Skumanich 1972). As the number of the fast rotators are few, and they show the severe dispersion in the acitivy–rotation relations, the saturation region is not as significant as that shown in the X-ray band. The results suggest that for slow rotators, the relations between rotation period and the Ca II $H$ and $K$ emissions as well as the photospheric variations could also be important probes of the physical dynamo process.

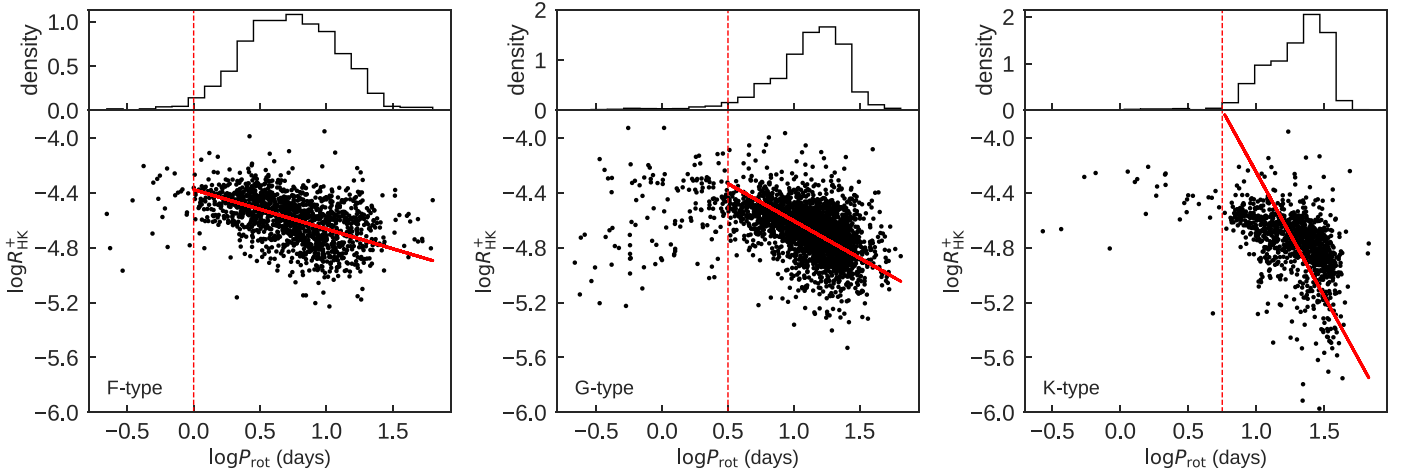

**Figure 6.** Relations between the $P_{\rm rot}$ and $R_{HK}^+$ indexes for F-type (left), G-type (middle), and K-type (right) stars. The top section of each panel shows the histogram of $\log P_{\rm rot}$. The red solid lines represent the fitted relations. The red dashed lines represent the lower limits of rotation periods for the fitting.

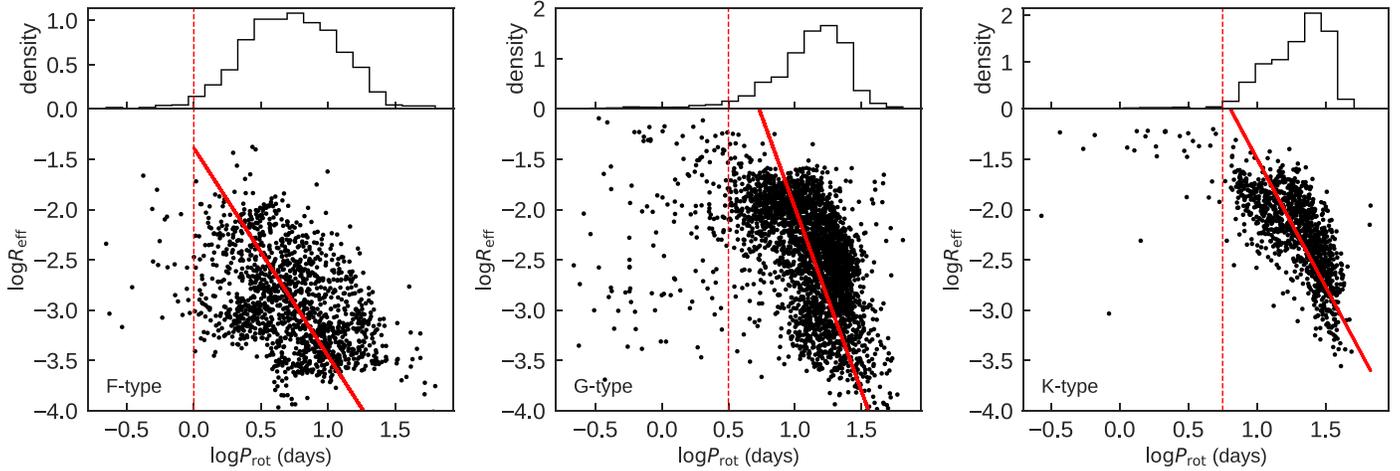

**Figure 7.** Relations between the $P_{\rm rot}$ and $R_{\rm eff}$ indexes for F-type (left), G-type (middle), and K-type (right) stars. The top section of each panel shows the histogram of $\log P_{\rm rot}$. The red solid lines represent the fitted relations. The red dashed lines represent the lower limits of rotation periods for the fitting.

**Table 3**
Power-law Index of Activity–Rotation Relations Fit for Different Type Stars

| Parameter | Relation | F-type | G-type | K-type |
|---|---|---|---|---|
| $\beta$ | $R_{HK}^{+}$–$P_{\rm rot}$ | $-0.29 \pm 0.01$ | $-0.54 \pm 0.01$ | $-1.81 \pm 0.05$ |
|  | $R_{\rm eff}$–$P_{\rm rot}$ | $-2.07 \pm 0.01$ | $-3.64 \pm 0.01$ | $-2.55 \pm 0.01$ |

## 5. Conclusion

We constructed two updated largest catalogs with stellar activity proxies in the LAMOST–*Kepler* program. One contains 59,816 F-, G-, and K-type stars with the chromospheric activity proxies $S$ and $R_{HK}^{+}$. Another one includes 5575 stars with the photospheric activity proxy $R_{\rm eff}$ and the rotation periods. We studied the relations between the activity proxies, as well as the relations between activity proxies and rotation periods.

The $\log R_{HK}^{+}$ shows positive correlation with $\log R_{\rm eff}$. There exists a power-law relation between $R_{HK}^{+}$ and $R_{\rm eff}$. The power-law indexes fitted for F- and G-type stars are 0.20 and 0.18, while the value fitted for K-type stars is 0.46.

Our analysis confirmed the relations between the two activity indices ($R_{HK}^{+}$ and $R_{\rm eff}$) and rotation period. For stars with rotation periods exceeding a few days, Ca II $H$ and $K$ emissions and photospheric variations generally decrease with increasing rotation period. The absolute values of the power-law index $\beta$ increase in the $R_{HK}^{+}$–$P_{\rm rot}$ relations from F- to K-type stars, while it does not show a similar trend in $R_{\rm eff}$–$P_{\rm rot}$ relations. Our results indicated that the relations between rotation period and the Ca II $H$ and $K$ emissions as well as the photospheric variations could also be important probes of the physical dynamo process.

The authors would like to thank the referees for the constructive criticism and useful advice, which helped us greatly improve the paper. The authors acknowledge support from the Joint Research Fund in Astronomy (U1631236) under cooperative agreement between the National Natural Science Foundation of China (NSFC) and Chinese Academy of Sciences (CAS), and grants 11522325, and 11873023 from the National Natural Science Foundation of China, and the Fundamental Research Funds for the Central Universities and Youth Scholars Program of Beijing Normal University. T.L. acknowledges funding from an Australian Research Council DP grant DP150104667, the Danish National Research Foundation (grant DNRF106). H.H. acknowledges the support of the Astronomical Big Data Joint Research Center, cofounded by the National Astronomical Observatories, Chinese Academy of Sciences and the Alibaba Cloud. This paper includes data collected by the *Kepler* Discovery Mission, whose funding is provided by NASA's Science Mission Directorate. Guoshoujing Telescope (the Large Sky Area Multi-Object Fiber Spectroscopic Telescope, LAMOST) is a National Major Scientific Project built by the Chinese Academy of Sciences. Funding for the project has been provided by the National Development and Reform Commission. LAMOST is operated and managed by the National Astronomical Observatories, Chinese Academy of Sciences.

### ORCID iDs

Shaolan Bi https://orcid.org/0000-0002-7642-7583
Jie Jiang https://orcid.org/0000-0001-5002-0577
Han He https://orcid.org/0000-0001-9352-9189
Zhishuai Ge https://orcid.org/0000-0002-2614-5959